\documentclass[reprint, superscriptaddress,
 amsmath,amssymb,
 aps, 
pra,
floatfix,
]{revtex4-2}

\usepackage{graphicx}% Include figure files
\usepackage{dcolumn}% Align table columns on decimal point
\usepackage{bm}% bold math
\usepackage{braket}
\usepackage{pifont}
\usepackage{amsmath}
\usepackage{array}
\begin{document}
\preprint{APS/123-QED}

\title{Optimization Framework for Full-Skyrmionic Polarimetry}

\author{Luo Wang}
\affiliation{School of Optics and Photonics, Beijing Institute of Technology, Beijing, China.}
\affiliation{Key Laboratory of Photonics Information Technology, Ministry of Industry and Information Technology, Beijing 100081, China.}
\affiliation{National Key Laboratory on Near-surface Detection, Beijing 100072, China.}

\author{Haiyang Zhang}
\email{ocean@bit.edu.cn}
\affiliation{School of Optics and Photonics, Beijing Institute of Technology, Beijing, China.}
\affiliation{Key Laboratory of Photonics Information Technology, Ministry of Industry and Information Technology, Beijing 100081, China.}
\affiliation{National Key Laboratory on Near-surface Detection, Beijing 100072, China.}

\author{Zilong Zhang}
\email{zlzhang@bit.edu.cn}
\affiliation{School of Optics and Photonics, Beijing Institute of Technology, Beijing, China.}
\affiliation{Key Laboratory of Photonics Information Technology, Ministry of Industry and Information Technology, Beijing 100081, China.}
\affiliation{National Key Laboratory on Near-surface Detection, Beijing 100072, China.}

\author{Hanqiu Zhang}
\affiliation{School of Optics and Photonics, Beijing Institute of Technology, Beijing, China.}
\affiliation{Key Laboratory of Photonics Information Technology, Ministry of Industry and Information Technology, Beijing 100081, China.}
\affiliation{National Key Laboratory on Near-surface Detection, Beijing 100072, China.}

\author{Changming Zhao}
\affiliation{School of Optics and Photonics, Beijing Institute of Technology, Beijing, China.}
\affiliation{Key Laboratory of Photonics Information Technology, Ministry of Industry and Information Technology, Beijing 100081, China.}
\affiliation{National Key Laboratory on Near-surface Detection, Beijing 100072, China.}

\date{\today}

\begin{abstract}
    Optical skyrmionic beams hold considerable potential for applications in quantum information and weak-intensity detection. These applications require accurate measurements of full topological observables using imaging Stokes polarimetry, with measurement states optimized in advance to limit the effect of stochastic error sources. However, the optimization framework for general Stokes polarimetries is based on a linear reconstruction chain for Stokes vectors and cannot quantitatively evaluate the nonlinear reconstruction of topological observables. Here, we define full-skyrmionic polarimetry (FSP) and construct its optimization framework from the reconstruction chain for full topological observables. Based on existing polarization evaluating functions (PEF), we introduce topological evaluating functions (TEF) and develop a multi-stage FSP optimization methodology. Numerical results confirm that topological optimization can further improve the reconstruction of topological observables, even when the measurement states are already optimal for polarization. Increasing the redundant projection number also effectively suppresses the effects of all stochastic error sources, providing a low-cost methodology to improve the reconstruction of both polarization and topological observables. The FSP optimization framework can identify topologically optimal configurations for arbitrary FSP structures, providing a guideline for high-precision full-skyrmionic measurements at weak-intensity levels and potential skyrmion ellipsometry.
\end{abstract}

\maketitle
%\tableofcontents
\section{Introduction}
    Optical skyrmionic beams are topologically protected vector beams whose topology is encoded in Stokes pseudospin fields $\bm{n}=\left\{s_1,s_2,s_3\right\}$. They show great potential in photonic entanglement \cite{Ornelas2024,Kam2026}, topological information transmission \cite{Peters2026,Wang2024a,Guo2026}, single-photon structured light sources \cite{Kam2026,Ma2025,Liu2026}, and novel light-matter interactions \cite{Ma2025,Mitra2025,Bo2026} due to their inherent topological number channels and robustness against perturbations. The topology of Stokes skyrmions is manifested in the fact that, across the propagation plane, the pseudo-spin field $\bm{n}(x,y)$ can always wrap around the entire Poincar\'e sphere an integer number of times \cite{Shen2025}, which is known as the skyrmion number $N_{sk}$. This topological number is defined by the area integral of the local topological density $q_{sk}$ and is expected to remain stable under weak perturbations.
    
    Currently, it is widely known that Skyrmionic beams can be generated by superposing two OAM beams with mutually orthogonal polarization states \cite{SrinivasaRao2024,Cisowski2023,Ye2024}. The relative components and cavities, such as metasurfaces, nano-quantum emitters (QEs), metafibers, interferometers, and Sagnac cavities, have been reported recently \cite{Liu2026,Cheng2026,He2024,Saito2025,Shen2022,Guo2025}. However, compared to the diverse generation methodologies, the high-precision measurement of full-skyrmionic observables has not received sufficient attention. Despite the fact that the latter is equally important in advancing Skyrmionic quantum and weak-intensity applications, which demand the optimization of measurement configurations to constrain the effects of stochastic error sources and maximize measurement quality. 
    
    The full-skyrmionic measurement represents the measurement of the complete set of observables of a skyrmionic beam, including both the general polarization textures and unique topological properties. The complete reconstruction chain naturally involves multiple stages. For high-precision full-skyrmionic measurements at weak-intensity levels, imaging Stokes polarimetry is the main candidate since it provides complete polarization observables and possesses a mature optimization framework. This technique modulates and projects the incident Stokes vector field onto a set of analysis states using a polarization state analyzer (PSA) and then linearly reconstructs the Stokes vector field \cite{Foreman2015,BalthasarMueller2016,Chipman2018,AlKhafaji2022}. The topological and other observables are further extracted from the reconstructed vector fields. Therefore, it can also be called full-skyrmionic polarimetry (FSP). However, existing optimization frameworks for Stokes polarimetry, including evaluating functions and optimization methodology, are based on the linear reconstruction chain of the Stokes field and cannot quantitatively evaluate the measurement quality of topological observables, which results in most skyrmions experiments still using the simplified PSA \cite{Shen2022,Guo2025,Zhang2025,Yang2026} configuration and lacking metrics to quantify the accuracy and stability. This could make the measurements sensitive to environment and potential positional offsets of optical elements, and introduce extra amplification of stochastic noise. At weak-intensity levels, the limitations may predictably cause severe degradation of measured skyrmionic observables. Thus, one of the key theoretical challenges in high-precision FSP is to investigate the multi-stage reconstruction chain of skyrmionic observables and develop a new optimization framework for FSP.

    In this article, we theoretically develop the optimization framework for full-skyrmionic polarimetry (FSP) from the reconstruction chain of skyrmionic observables. First, we investigate the effects of two main stochastic error sources, analysis matrix mismatch and composite stochastic noises, on the measurements of FSP, and introduce key topological evaluating functions (TEF) to quantitatively evaluate the reconstruction quality of topological observables (topological density and skyrmion number). Then, we develop a multi-stage optimization methodology that combines polarization and topological optimization. The numerical results show that the topological optimization can further improve the topological reconstruction quality, even after the measurement states are already optimal for polarization. Besides, increasing the redundant projection numbers is verified to reduce the effects of both stochastic error sources on all skyrmionic observables, offering an extra low-cost methodology to optimize FSP. This work directly connects FSP optimization to the reconstruction quality of skyrmionic observables, which provides a guideline for high-precision full-skyrmionic measurements at weak-intensity levels and potential skyrmion ellipsometry. 
    
\section{Basic concepts}
    \begin{figure*}[htb]
    \centering
    \includegraphics[width = 18.0cm]{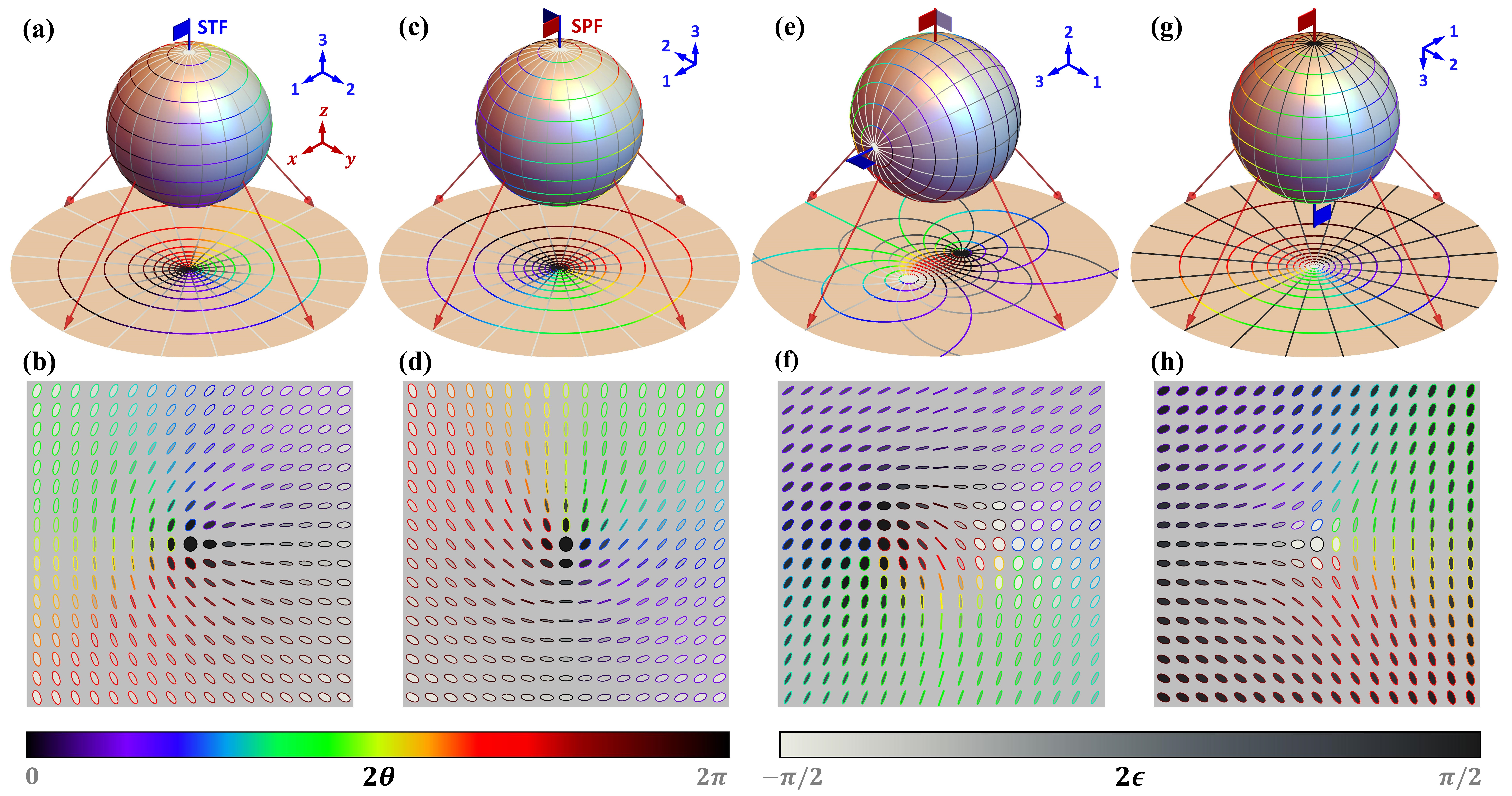}
       \caption{Various textures within a same Skyrmionic family ($l_1=0$, $l_2=1$), where the first row visualizes the stereographic projections between $\mathbb{R}^2$ and $\mathbb{S}^2$, where the relative position of the red SPF and blue STF determine the texture type. The second row visualizes the distribution of polarization ellipses corresponding to each texture. The rainbow-colored bars represent the double azimuth of the polarization ellipse, while the black \& white bars represent the double ellipticity. (a)$\sim$(b) N\'eel type (basic type), SPF and STF are completely coincident. (c)$\sim$(d) Bloch type, SPF and STF are coaxial but not coincident. (g)$\sim$(h) Anti-type, SPF and STF are completely opposite.}
       \label{fig:01}
       \vspace{-1.0\baselineskip}
    \end{figure*}
    
    \subsection{Skyrmionic textures}
        In the theory of paraxial Stokes skyrmions, a spatially distributed Skyrmionic state-vector or a Skyrmionic beam in pure-state can be constructed by superposing two mutually orthogonal eigenstates $\ket{\uparrow}$ and $\ket{\downarrow}$ \cite{Gao2020,SrinivasaRao2024,Cisowski2023,Ye2024}, along with their respective vortex modes $E_{l_1}(r,\phi)$ and $E_{l_2}(r,\phi)$ carrying different orbital angular momentum (OAM), where the vortex mode components can be Laguerre-Gaussian (LG) or Bessel modes:  
        \begin{equation}
            \ket{\Psi(r,\varphi)}=E_{l_1}(r,\varphi)\ket{\downarrow}
            +E_{l_2}(r,\varphi)\ket{\uparrow}
        \end{equation}

        The mutually orthogonal eigenstates $\{\ket{\downarrow}, \ket{\uparrow}\}$ correspond to a pair of coordinate points in opposite directions on the Poincar\'e sphere. The most common combination is the left-hand and right-hand circular polarization states $\{\ket{L}, \ket{R}\}$, which correspond to the south and north poles. Here, a classic case involves employing the LG vortex mode in Eq.(2) with $l_1=0$ and $l_2=1$. Substituting them into Eq.(1) will generate a N\'eel type Skyrmionic texture. The corresponding numerical results are presented in FIG.1-(a) and (b).
        \begin{equation}
            E_{l}(r,\varphi)=A_{l}\left( \frac{\sqrt{2}r}{\omega_0} \right)^{|l|}
            \bm{L}_p^{|l|} \left( \frac{2r_0^2}{\omega_0^2} \right)
            e^{-\frac{r_0^2}{\omega_0^2}}e^{-il\varphi}
        \end{equation}

        In fact, arbitrary Skyrmionic textures with a fixed Skyrmion number can be achieved by performing a specific unitary transformation,$\{ \ket{\downarrow},\ket{\uparrow} \}=\mathbf{U}\{ \ket{L},\ket{R} \}$ ($\mathbf{U}\in \mathrm{SU}(2)$), on the eigenstate of a N\'eel type (basic type) texture \cite{Cisowski2023,Ye2024}. A whole set of textures generated from these $\mathrm{SU}(2)$ transformations constructs a skyrmionic family. Furthermore, each $\mathrm{SU}(2)$ transformation of the eigenstate can be uniquely mapped to a $\mathrm{SO}(3)$ rotation in $\mathbb{R}^3$ space, which implies that we can define two "flags" with unit length and classify texture types within the same family based on the relative position of the flagpole and flag surface.
        
        As shown in the first row of FIG.1, the red flag fixed on the $S^2-$sphere is defined as the stereographic projection flag (SPF). The negative direction of the flagpole represents the direction of the stereographic projection, $S^2\mapsto\mathbb{R}^2\cup \{\infty\}$. The $x-y-z$ reference frame is tied to the SPF, while the plane $x-y$ is the actual propagation plane of skyrmionic beams. The blue flag is defined as the skyrmionic texture flag (STF). The $1-2-3$ reference system reflects the orientation of the $S^2-$sphere wrapped by the Skyrmionic texture in $\mathbb{R}^3$ abstract space. When the STF and SPF completely overlap, as shown in (a) and (b), the Skyrmionic texture exhibits a basic N\'eel type; When the STF and SPF have the same flagpole but differ in flag surface orientation by $\pm\pi/2$, as shown in (c) and (d), the texture presents the Bloch types; When the STF and SPF have a $\pi/2$ angle between their flagpoles, as shown in (e) and (f), the texture appears as bimeron types with two $\mathrm{C}-$points. When the STF and SPF have opposite flagpoles, as shown in (g) and (h), the texture appears as an anti-Skyrmionic type.
        
    \subsection{Polarization and topological observables}
        The polarization and topological observables that the FSP reconstructs in a single measurement include the pseudospin field $\bm{n}(x,y)$, topological density $q_{sk}(x,y)$, and skyrmion number $N_{sk}$. These physical quantities are all nonlinear functions of Stokes parameters $S_0\sim S_3$. In paraxial cases \cite{Bliokh2015}, four Stokes parameters can be mapped from three Pauli matrices and an identity matrix to analytical forms: $\bm{\sigma}_j\mapsto S_j=\bra{\Psi} \bm{\sigma}_j \ket{\Psi}$ ($j=0,\cdots,3$) \cite{Shen2024,Ye2024}, and further expanded as follows: 
        \begin{equation}
            \begin{aligned}
                \begin{gathered}
                    S_0=\langle \Psi_1^{}\Psi_1^{*}+\Psi_2^{*}\Psi_2^{} \rangle,\ 
                    S_1=\langle \Psi_1^{}\Psi_1^{*}-\Psi_2^{*}\Psi_2^{} \rangle \\
                    S_2=\langle \Psi_1^{}\Psi_2^{*}+\Psi_1^{*}\Psi_2^{} \rangle,\ 
                    S_3=i\langle \Psi_1^{}\Psi_2^{*}-\Psi_1^{*}\Psi_2^{} \rangle
                \end{gathered}
            \end{aligned}
        \end{equation}

        Since only the last three Stokes parameters carry polarization information, we define a polarization vector, $\bm{v}=\{S_1,S_2,S_3\}$ for convenience. The pseudospin $\bm{n}$ can be naturally written as its normalization: $\bm{n}=\bm{v}/\|\bm{v}\|=\{s_1,s_2,s_3\}$, where the three elements are precisely equivalent to the parametric equations for the Poincar\'e sphere: $s_1= \cos(2\epsilon)\cos(2\theta)$, $s_2=\cos(2\epsilon)\sin(2\theta)$, and $s_3 = \sin(2\epsilon)$. The angular parameters $\theta$ and $\epsilon$ correspond to the azimuth and ellipticity of the polarization ellipse, respectively (the factor of 2 arises from the double covering of $\mathrm{SO}(3)$ by $\mathrm{SU}(2)$ \cite{Goldberg2021}). Therefore, the polarization ellipse fields in the second row of FIG.1 can be naturally derived from the pseudospin texture:
        \begin{equation}
            \theta(x,y)=\frac{1}{2} \arctan \left( \frac{s_2}{s_1} \right),\ 
            \epsilon(x,y)=\frac{1}{2} \arcsin\left(s_3\right)
        \end{equation}

        Here, $\theta$ and $\epsilon$ are restricted to $(0,\pi]$ and $[-\pi/4,\pi/4]$, respectively, to ensure the uniqueness of the mapping. The polarization ellipse field directly visualizes the skyrmionic texture in physical space. Spatial variations in ellipse shape and chirality can be used to identify polarization singularities \cite{Bauer2016,MataCervera2025}. In addition to these spatial polarization properties, skyrmionic beams are also characterized by local and global topological observables. In the propagation plane $x-y$, the local topology is described by the topological density $q_{sk}(x,y)$, which is defined in terms of the pseudospin field and its derivatives with respect to $x$ and $y$ \cite{Shen2025,Wang2024b,Wu2025}:
        \begin{equation}
            q_{sk}(x,y)=\frac{1}{4\pi} \bm{n}\cdot\left( 
            \frac{\partial\bm{n}}{\partial x} \times 
            \frac{\partial\bm{n}}{\partial y}
            \right)
        \end{equation}

        This quantity describes the local contribution to the topological mapping. Integrating it over the selected effective region $\bm{\Omega}$ can provide the skyrmion number:        
        \begin{equation}
            N_{sk}=\iint_{\bm{\Omega}} q_{sk}(x,y) dxdy
        \end{equation}

        The skyrmion number evaluates the number of times the pseudospin field wraps around the Poincar\'e sphere and is known as a fundamental global topological observable. Other global observables include the polarity $p$, vorticity $m$, and helicity $\gamma$ \cite{Shen2022,Shen2024,Yang2025}, where the polarity $p$ describes the relative longitudinal orientation of the core and background pseudospins. The vorticity $m$ counts the number of times the pseudospin rotates as one moves around the core. The helicity $\gamma$ provides the angular offset of the pseudospin from the radial direction, distinguishing N\'eel, Bloch, and other intermediate textures. These quantities describe specific features of the texture but cannot replace the skyrmion number. Here, we define the Stokes vector $\bm{S}(x,y)$ and the pseudospin field $\bm{n}(x,y)$ as the reconstructed polarization observables; the topological density $q_{sk}$ and the Skyrmion number $N_{sk}$ as the final reconstructed topological observables for FSPs, which generate a nonlinear reconstruction chain starting from the detected intensity images: $\{I_1,\cdots, I_N\}\to \{\bm{S}(x,y), \bm{n}(x,y)\}\to \{q_{sk}(x,y), N_{sk}\}$.

\section{Optimization framework}
    \subsection{Projection and reconstruction model}
        \begin{figure*}[htb]
    	\centering
        \includegraphics[width = 17.65cm]{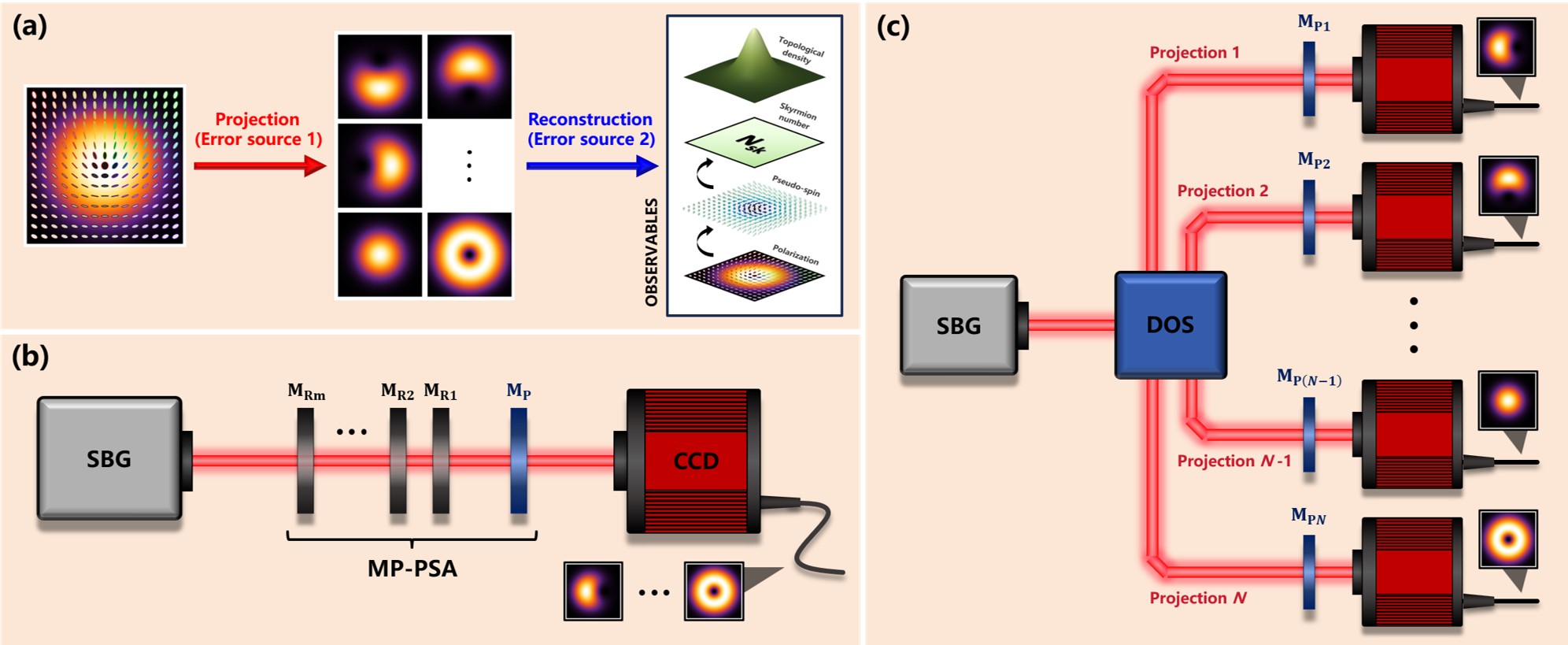}
           \caption{Diagrams of two types of Full-Skyrmionic polarimetry (FSP). SBG: Skyrmionic beam generator. (a) Polarization projections and the reconstruction of polarization and topological observables, where error sources 1 and 2 affect the projection and reconstruction, respectively. (b) Division-of-time FSP (DOT-FSP), which requires sequential acquisition but allows a large number of projections, making it a preferred choice for high-precision skyrmionic measurements. (c) Single-shot division-of-space FSP (DOS-FSP), which acquires intensity images simultaneously through parallel polarization projections. The projection number is typically limited to $N=4$ since increasing the projection requires additional optical channels.}
           \label{fig:02}
           \vspace{-1.0\baselineskip}
        \end{figure*}
        
        As shown in FIG.2, the construction logic of FSP is analogous to that of full-Stokes polarimetry \cite{Chipman2018,Dong2020}. It requires recording intensity images $I_1(x,y),\ldots,I_N(x,y)$ through polarization projections performed sequentially in time or simultaneously in separate optical channels. The division-of-time FSP (DOT-FSP) in FIG.2-(b) uses a single CCD camera and a multi-plate polarization state analyzer (MP-PSA) composed of phase retarders (PR) and a linear polarizer (LP). The PSA is adjusted sequentially to provide $N\ge 4$ analysis states, and the camera records the intensity image for each state. The division-of-space FSP (DOS-FSP) in FIG.2-(c) uses $N$ cameras to record the intensity images simultaneously through $N$ separate optical channels, each with its own analysis state. Technically, DOS-FSP can achieve single-shot skyrmionic measurements, but increasing the projections requires additional optical channels and cameras, so the projection number is typically limited to $N=4$. In contrast, DOT-FSP has fewer components and stochastic noise sources, and it facilitates a large-scale increase in the projection numbers, which makes it better suited for high-precision skyrmionic measurements. Thus, we focus on DOT-FSP in the following analysis.

        The polarization projection of an FSP is described as a linear mapping of the Stokes vector in Mueller calculus, given by $I_i = \mathbf{a}_i(\mathbf{M}_i)\bm{S}$, where the vector $\mathbf{a}_i$ is the projection basis and is defined by the first row of the $4\times 4$ Mueller matrix $\mathbf{M}_i$ for the $i-$th projection ($1\le i\le N$) \cite{Foreman2015,Chipman2018,Teng2023,deMelloKoch2025}. For a MP-PSA in FIG.2-(b), the total Mueller matrix can be described as the sequential product of the Mueller matrices of the individual components:
        \begin{equation}
            \mathbf{M}_i\left( \beta_P,\beta_{R1},\cdots,\beta_{Rm} \right)=
            \mathbf{M}_P\left( \beta_P \right) \prod_{j=1}^{m} \mathbf{M}_{Rj} 
            \left( \beta_{Rj} \right)
        \end{equation}

        Here, $\mathbf{M}_P$ and $\mathbf{M}_{Rj}$ represent the Mueller matrices of the LP and the $j$-th PR (counting from right to left), respectively, where the variables include $m+1$ azimuth angles $\beta_P$ and $\beta_{R1},\ldots,\beta_{Rm}$. The detailed derivation of these Mueller matrices is provided in \textbf{Appendix A}. Together, the $N$ projection vectors $\bm{a}_1,\cdots,\bm{a}_N$ form the real $N\times 4$ analysis matrix $\mathbf{A}$. The complete polarization projection is written as $\bm{I}(x,y)=\mathbf{A}\bm{S}(x,y)$, where $\bm{I}=\{I_1,\cdots,I_N\}$ is the set of detected intensity images. The Stokes vector field is linearly reconstructed as $\bm{S}_R(x,y)=\mathbf{A}^{+}\bm{I}(x,y)$, where $\mathbf{A}^{+}$ is the Moore–Penrose pseudoinverse of $\mathbf{A}$. When $N=4$, and $\mathbf{A}$ is nonsingular, $\mathbf{A}^{+}=\mathbf{A}^{-1}$. The pseudospin field and topological observables can be reconstructed from $\bm{S}_R(x,y)$.
        
    \subsection{Errors sources and polarization evaluating functions}
        Similar to conventional Stokes polarimetry, different stochastic error sources will arise at different stages of each measurement in FSPs. These error sources propagate through the nonlinear reconstruction chain, degrading polarization and topological observables to different degrees. Therefore, the high-precision FSP at the weak-intensity level requires modeling these error sources and defining suitable evaluating functions to quantify measurement errors and guide their optimization. We examine these stochastic error sources in detail below.
        
        \vspace{0.4em}
        \noindent
        $\blacktriangleright$\enspace
        \textbf{Error Source 1 (Analysis matrix mismatch):}
        \vspace{0.4em}
        
        As shown in FIG.2-(a), this error source occurs at the projection stage and arises from a systematic mismatch between the actual PSA analysis matrix $\mathbf{A}_r$ and the calibrated matrix $\mathbf{A}_c$ used for reconstruction. The potential causes include random positional offsets of optical elements, residual calibration errors, operating wavelength shifts, and possible mechanical deformations \cite{SuarezBermejo2019,Wu2022}. The matrix mismatch remains fixed throughout each measurement. When error source 1 exists, the projection can be rewritten in the following form:
        \begin{equation}
            \bm{I}(x,y)=\mathbf{A}_r\cdot \bm{S}_I(x,y)
            =\left( \mathbf{A}_c+\Delta\mathbf{A} \right)
            \cdot \bm{S}_I(x,y)
        \end{equation}

        It should be emphasized that, since the analysis matrix is a complex function derived from Mueller calculus, the specific mismatch cannot simply be added or subtracted linearly. $\Delta \mathbf{A}$ is used solely for conceptual description.

        \vspace{0.4em}
        \noindent
        $\blacktriangleright$\enspace
        \textbf{Error Source 2 (Composite stochastic noise):}
        \vspace{0.4em}

        This error source occurs at the reconstruction stage shown in FIG.2-(a). The detectors in arbitrary Stokes polarimetry inevitably suffer from various stochastic noise sources, such as thermal noise, dark noise, and shot noise, which follow different statistical distributions. The stochastic noises can be modeled as an additive term, $\delta \bm{I} (x,y)$, to the projected intensity images. Then, the noise propagates through the nonlinear reconstruction chain and can be amplified in the reconstructed observables \cite{Goudail2016a,Goudail2016b,Mu2019}. With error source 2 included, the reconstruction equation is rewritten as:
        \begin{equation}
            \bm{S}_R(x,y)=\mathbf{A}_c^+ \cdot 
            \left[ \bm{I}(x,y)+\delta \bm{I}(x,y) \right]
        \end{equation}

        Although some simplified models treat $\delta\bm{I}$ as Gaussian noise, we use a composite stochastic noise model, $\delta\bm{I}(x,y)=\bm{n}_g(x,y)+\bm{n}_p(x,y)$, that combines Gaussian noise with intensity-dependent Poisson noise to better match experimental situations. Substituting Eq.(8) into Eq.(9) can provide the reconstructed Stokes field: $\bm{S}_R(x,y)=\mathbf{A}_c^+\mathbf{A}_r\bm{S}_I(x,y)+\mathbf{A}_c^+\delta \bm{I}(x,y)$. Here, we further define the absolute Stokes error as $\Delta \bm{S}(x,y)=\bm{S}_R-\bm{S}_I$, and then express the relative Stokes error using the Hilbert-Schmidt norm as $\|\Delta \bm{S}\|/\|\bm{S}_I\|$. Its supremum provides a global evaluation of the linear reconstruction quality of the Stokes field when both stochastic error sources exist. In order to relate this supremum to specific scalar evaluating functions, we introduce the triangle inequality and obtain the following inequality:
        \begin{equation}
            \sup \left( \frac{\|\Delta \bm{S}\|}{\bm{S}_I} \right) \leq 
            \|\mathbf{B}\|+\|\mathbf{A}_c^+\| \cdot 
            \sup \left( \frac{\|\delta \bm{I}\|}{\bm{S}_I} \right)
        \end{equation}

        In Eq.(10), we define the $4\times4$ matrix $\mathbf{B}=\mathbf{A}_c^{+}\mathbf{A}_r-\mathbf{I}_4$ to characterize the reconstruction error caused by Error Source 1. This equation relates the upper bound on the relative Stokes error to two separate metrics, rather than to a single metric of condition number (CN). The first metric, $\|\mathbf{B}\|$, describes the effect of error source 1 and remains fixed for each measurement. The second metric, $\|\mathbf{A}_c^{+}\|$, describes the amplification of composite stochastic noise from Error Source 2 and depends only on the PSA configuration. Its square equals the equally weighted variance (EWV) introduced by Sabatke et al \cite{Sabatke2000,Goudail2016b}. Therefore, we define two polarization evaluating functions (PEF), $\mathcal{F}_1=\|\mathbf{B}\|$ and $\mathcal{F}_2=\|\mathbf{A}_c^{+}\|$, which can be further expanded as follows:
        \begin{equation}
            \mathcal{F}_1=\sqrt{\mathrm{Tr} \left[ \mathbf{B}^\dagger\mathbf{B} \right]},\ 
            \mathcal{F}_2=\sqrt{\mathrm{Tr} \left[ (\mathbf{A}_c^+)^\dagger
            (\mathbf{A}_c^+)\right]}
        \end{equation}

    \subsection{Topological evaluating functions}
        \begin{figure*}[htb]
    	\centering
        \includegraphics[width = 17.6cm]{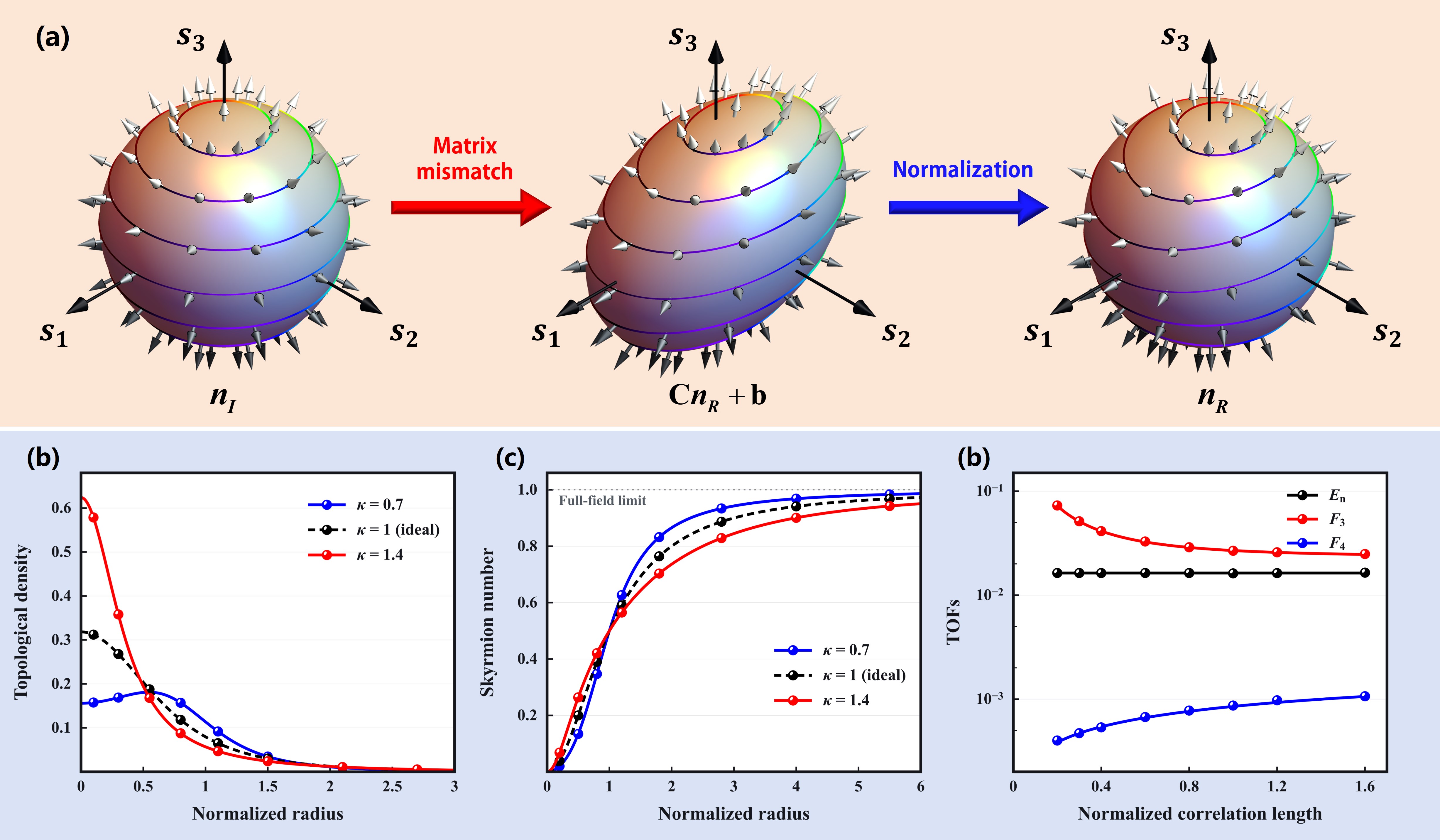}
           \caption{(a)$\sim$(c) Effects of analysis matrix mismatch (Error Source 1) on the reconstruction of the pseudospin and topological observables. (a) The incident pseudospin $\bm{n}_I$ are mapped to the reconstructed pseudospin $\bm{n}_R$ through the affine transformation $\mathbf{C}\bm{n}_I+\mathbf{b}$ followed by normalization. The surfaces show the distortion of the full set of pseudospins. (b) Radial profiles of the reconstructed topological density under anisotropic mismatch, with $\mathbf{C}=\operatorname{diag}(\kappa,\kappa,1)$, $\bm{b}=0$, and $\kappa=0.7,1,1.4$ ($\kappa=1$ corresponds to ideal reconstruction). (c) Reconstructed skyrmion number as a function of the normalized radius $R=r/r_0$, where all curves approach the full-field limitation of 1 (dashed line). (d) Dependence of $\mathcal{F}_3$, $\mathcal{F}_4$, and $E_n$ on the noise correlation length, evaluated over a circular integration region of radius $4r_0$, with the fixed 2\% relative RMS perturbation of $\bm{v}_R$.}
           \label{fig:03}
           \vspace{-1.0\baselineskip}
        \end{figure*}
        
        The PEFs $\mathcal{F}_1$ and $\mathcal{F}_2$ apply to both single-pixel and imaging Stokes polarimetry. They evaluate the linear reconstruction of the Stokes vector. However, FSP further requires the extraction of the topological density $q_{sk}(x,y)$ and the Skyrmion number $N_{sk}$. These topological observables depend on the normalization of the pseudospin field $\bm{n}(x,y)$, its spatial derivatives, and the definition of the effective integration region. The reconstruction quality of these observables cannot be adequately assessed based solely on the relative Stokes error and the PEFs. Therefore, this subsection introduces topological evaluating functions (TEF) to assess the reconstruction of these topological observables. Note that when the scale of the polarization vector $\bm{v}_R$ approaches the noise level, the relevant integration regions must be evaluated separately. The principle is shown in \textbf{Appendix B}.
        \begin{equation}
            \mathbf{A}_c^+\mathbf{A}_r=\mathbf{I}_4+\mathbf{B}=
            \begingroup
                    % ===== 列间距：相邻两列的留白 = 2 × \arraycolsep =====
                    \setlength{\arraycolsep}{5pt}  % 此处对应 12pt 的列间留白
                    % ===== 行间距：基线上方高度 + 基线下方深度 =====
                    \renewcommand{\arraystretch}{1}
                    \setlength{\extrarowheight}{0pt}
                    \setbox\strutbox=
                    \hbox{
                    \vrule height 11pt depth 5pt width 0pt
                    }
            \begin{pmatrix}
                k_{00} & \mathbf{k}^T \\
                \mathbf{b} & \mathbf{C}
            \end{pmatrix}
            \endgroup
        \end{equation}

        We first investigate how Error Source 1 degrades the reconstruction of topological observables. The reconstruction equation is written as $\bm{S}_R(x,y)=\mathbf{A}_c^{+}\mathbf{A}_r\bm{S}_I(x,y)$. Here, the matrix product $\mathbf{A}_c^{+}\mathbf{A}_r$ can be expressed in the block form shown in Eq.(12). $\mathbf{C}$ is a $3\times3$ matrix describing the coupling among the three polarization components, and $\bm{b}$ is a three-component vector representing crosstalk from intensity into these components. Then, the reconstructed pseudospin field $\bm{n}_R(x,y)$ can be written as a spherical map in Eq.(13).
        \begin{equation}
            \bm{n}_R=\frac{\mathbf{C}\bm{n}_I+\mathbf{b}}{\|\mathbf{C}\bm{n}_I+\mathbf{b}\|}
        \end{equation}
        
        For each complete measurement, Eqs.(12) and (13) describe a nonlinear mapping from the incident pseudospin field $\bm{n}_I(x,y)$ to the reconstructed field $\bm{n}_R(x,y)$, as illustrated in FIG.3-(a). When $\mathbf{C}$ is invertible and $\mathbf{C}\bm{n}_I+\mathbf{b}\neq0$, the resulting distortion in topological density is given analytically by $q_{skR}(x,y)=g(\bm{n}_I)q_{skI}(x,y)$, where the scaling factor $g(\bm{n}_I)$ is:
        \begin{equation}
            g(\bm{n}_I)=
            \frac{\mathbf{DET}(\mathbf{C})\left( 1+\bm{n}_I^T\mathbf{C}^{-1}\mathbf{b} \right)}{\|\mathbf{C}\bm{n}_I+\mathbf{b}\|^3}
        \end{equation}

        This equation establishes a direct connection between specific matrix mismatch and local topological density distortion. Its detailed derivation can be found in \textbf{Appendix B}. When $\mathbf{C}=a\mathbf{R}$ ($\mathbf{R}\in \mathrm{SO}(3)$) and $\mathbf{b}=0$, the scaling factor $g$ equals 1 (isotropic mismatch), and the uniform amplitude scaling and global rotation of the pseudospin leave the topological density unchanged. In other cases, arbitrary anisotropic mismatch will cause $g(\bm{n}_I)$ to vary with the pseudospin field, leading to spatially distributed distortions in $q_{skR}(x,y)$.

        However, due to topological protection, distortions in topological density do not equal the distortion of skyrmion number. For skyrmion fields that satisfy the compactification conditions \cite{Wang2024a}, the spherical map in Eq.(13) retains a mapping degree of 1 when $\mathbf{DET}(\mathbf{C})>0$ and $\|\mathbf{C}^{-1}\mathbf{b}\|<1$. In this case, the $\bm{n}_R$ sphere in FIG.3-(a) still encloses the origin and retains its orientation after the affine transformation $\mathbf{C}\bm{n}_R+\mathbf{b}$ and spherical mapping, so $N_{skR}=N_{skI}$. Numerical simulations in FIG.3-(b) show the radial distributions of the reconstructed topological density with different anisotropic mismatches. FIG.3-(c) shows the evolution of reconstructed skyrmion numbers with the normalized integration radius. For all cases shown, the reconstructed skyrmion number approaches the ideal value of 1 as the integration region expands to cover the full field (full-field limit), which also matches the topological protection of skyrmions under appropriate boundary conditions \cite{Wang2024a,Wang2026b}. However, for a fixed integration window, matrix mismatch can still change the topological density distribution and introduce additional truncation errors in the reconstructed skyrmion number. Therefore, the metric of the skyrmion number alone is not enough to fully evaluate reconstruction quality, since it can miss distortions in local topological observables. Here, we define two new topological evaluating functions (TEF), $\mathcal{F}_3$ and $\mathcal{F}_4$, to quantify the reconstruction quality of local and global topological observables, respectively.
        \begin{equation}
            \mathcal{F}_3=\sqrt{
            \frac{\mathbf{E}\left\{ \iint_{\bm{\Omega}} [q_{skR}(x,y)-q_{skI}(x,y)]^2 dxdy \right\}}
            { \iint_{\bm{\Omega}}q_{skI}^2(x,y)dxdy }
            }
        \end{equation}

        \begin{equation}
            \mathcal{F}_4=\sqrt{\mathbf{E}\left[ (N_{skR}-N_{skI})^2 \right]}
        \end{equation}

        $\mathbf{E}$ denotes the mathematical expectation, which generally covers the average influence of stochastic noise sources when the matrix mismatch is held constant. $\mathcal{F}_3$ measures the relative RMS error in the local topological density, while $\mathcal{F}_4$ measures the absolute RMS error in the skyrmion number over the same effective integration region. In numerical simulations, the incident and reconstructed skyrmion fields must be evaluated on the same sampling grid and over the same effective integration region to avoid introducing additional numerical errors.

        Then, to relate these TEFs directly to the two error sources, we start from the Stokes reconstruction error $\Delta\bm{S}=\mathbf{B}\bm{S}_I+\mathbf{A}_c^{+}\delta\bm{I}$, which is derived from Eqs.(8) and (9). For a nonzero polarization vector $\bm{v}_I$, a first-order Taylor expansion of $\bm{n}_I=\bm{v}_I/\|\bm{v}_I\|$ can provide $\delta\bm{n}=\mathbf{P}\delta\bm{v}/\|\bm{v}_I\|+\mathrm{O}(|\delta\bm{v}|^2/\|\bm{v}_I\|^2)$, where $\mathbf{P}=\mathbf{I}_3-\bm{n}_I\bm{n}_I^T$ projects the perturbation onto the tangent plane of the $\bm{n}_I$ sphere in FIG.3-(a). Thus, to first order, only the component of $\delta\bm{v}$ perpendicular to $\bm{n}_I$ changes the incident pseudospin, and this component is scaled by $1/\|\bm{v}_I\|$. Further performing a first-order expansion of the topological density can provide:

        \begin{equation}
            \begin{aligned}
                \begin{gathered}
                    \delta q_{sk}=\frac{1}{4\pi}
                    \left[
                    \delta\bm{n} \cdot \left( \frac{\partial\bm{n}}{\partial x} \times \frac{\partial\bm{n}}{\partial y} \right)+ \right.\\ \left.
                    \bm{n} \cdot \left( \frac{\partial \delta\bm{n}}{\partial x} \times \frac{\partial \bm{n}}{\partial y} \right)+
                    \bm{n} \cdot \left( \frac{\partial \bm{n}}{\partial x} \times \frac{\partial \delta\bm{n}}{\partial y} \right)
                    \right] 
                    \\
                    \rightarrow 
                    \frac{1}{4\pi} \bm{n} \cdot \left[ 
                    \left( \frac{\partial \delta\bm{n}}{\partial x} \times 
                    \frac{\partial \bm{n}}{\partial y} \right)+
                    \left( \frac{\partial \bm{n}}{\partial x} \times 
                    \frac{\partial \delta\bm{n}}{\partial y} \right)
                    \right]
                \end{gathered}
            \end{aligned}
        \end{equation}

        Here, the pseudospin perturbation satisfies $\bm{n}\cdot\delta\bm{n}=0$. Since $\partial_x\bm{n}\times\partial_y\bm{n}$ is parallel to $\bm{n}$, the first term vanishes. The remaining two terms show that the topological density error depends on the pseudospin field as well as the spatial derivatives of both the field and its perturbation. Thus, two measurements with nearly equal Stokes error scales can still yield substantially different topological errors, as measured by $\mathcal{F}_3$ and $\mathcal{F}_4$.

        We now account for spatial sampling by the detector array. In practice, topological observables are reconstructed from a discrete grid of pixels. We stack the Stokes parameters at all pixels into a single column vector and define $\mathbf{J}_q$ as the Jacobian of the discrete topological density vector with respect to the Stokes vector. For an array of $M$ pixels, $\mathbf{J}_q$ is a real $M\times4M$ matrix that combines the effects of normalization and spatial differentiation. We denote the corresponding full-array forms of $\mathbf{B}$, $\mathbf{A}_c^{+}$, $\bm{S}_I$, and $\delta\bm{I}$ by $\mathbb{B}$, $\mathbb{A}$, $\mathbb{S}_I$, and $\delta\mathbb{I}$, respectively. For weak perturbations, the error in the discrete topological density is then given by:
        \begin{equation}
            \delta \bm{q} \approx \left\{ \delta q_1 \cdots \delta q_M \right\}
            =\mathbf{J}_q \left( \mathbb{B}\cdot\mathbb{S}_I
            +\mathbb{A}\cdot\delta\mathbb{I} \right)
        \end{equation}

        In Eq.(18), the two contributions to the topological density error arise from error sources 1 and 2, respectively. For a fixed matrix mismatch, assume that the detector noise has zero mean and covariance matrix $\mathbf{\Sigma}_I$. Taking the expectation and covariance of Eq.(18) can further provide the mean topological density error, $\bm{\mu}_q=\mathbf{E}(\delta\bm{q})\approx\mathbf{J}_q\mathbb{B}\cdot\mathbb{S}_I$, and its covariance matrix, $\mathbf{\Sigma}_q=\mathbf{Cov}(\delta\bm{q})\approx(\mathbf{J}_q\mathbb{A})\mathbf{\Sigma}_I(\mathbf{J}_q\mathbb{A})^T$. To express the TEFs $\mathcal{F}_3$ and $\mathcal{F}_4$ in terms of these statistics, we introduce the area integration weights $\bm{w}=(w_1,\cdots,w_M)^T$ and define $\mathbf{W}=\mathbf{diag}(\bm{w})$. In discrete form, the spatial integral of the squared topological density error is $\delta\bm{q}^T\mathbf{W}\delta\bm{q}$, while the error in the skyrmion number is $\bm{w}^T\delta\bm{q}$. Then, we write $\delta\bm{q}$ as the sum of its mean and a zero-mean fluctuation. Substituting this decomposition into the discrete forms of Eqs.(15) and (16) and eliminating the cross terms can provide:
        \begin{equation}
            \mathcal{F}_3^2=\frac{\bm{\mu}_q^T\mathbf{W}\bm{\mu}_q+
            \mathbf{Tr}(\mathbf{W}\cdot\mathbf{\Sigma_q})}
            {\bm{q}_I^T\mathbf{W}\bm{q}_I}
        \end{equation}

        \begin{equation}
            \mathcal{F}_4^2=(\bm{w}^T\bm{\mu}_q)^2+\bm{w}^T\mathbf{\Sigma}_q\bm{w}
        \end{equation}
        
        Substituting $\bm{\mu}_q$ and $\mathbf{\Sigma}_q$ into these expressions allows us to evaluate TEFs using the analysis matrices, incident Stokes vectors, and noise covariances. Since $\mathbf{W}$ is diagonal, the variance part of $\mathcal{F}_3$ sums the local topological density variances, whereas that of $\mathcal{F}_4$ also includes covariances between different pixels. Note that, even though the detector noise is independent across pixels, these covariances must also be included because derivatives at neighboring pixels may use the same sampling points. Besides, global $\mathrm{SO}(3)$ rotations of the pseudospin field can change the texture type within the same skyrmionic family while keeping $q_{sk}$ and $N_{sk}$ constant, as shown in Fig.1. We introduce an additional PEF, the pseudospin error $E_n$, for convenience in the following analysis.
        \begin{equation}
            E_n=\sqrt{ \frac{1}{\bm{\Omega}}\mathbf{E}
            \left[ \iint_{\bm{\Omega}} \| \bm{n}_R(x,y)-\bm{n}_I(x,y) \|^2 dxdy \right] }
        \end{equation}
        
        $E_n$ has a completely different meaning from $\{\mathcal{F}_3,\mathcal{F}_4\}$. FIG.3-(d) shows the evolution of these three evaluating functions as a function of the spatial correlation length $L_c$ at a fixed Stokes perturbation scale. The simulation uses a circular integration region of radius $4r_0$, where $r_0=\omega_0|A_0|/(\sqrt{2}|A_{\pm1}|)$ is the incident texture scale. Here, $L_c$ is the spatial correlation length of the effective Stokes perturbations (using the Gaussian correlation function $\rho(d)=\exp[-d^2/(2L_c^2)]$, where $L_c$ is defined as the distance at which the correlation coefficient falls to $e^{-1/2}$ \cite{Rasmussen2006}). As $L_c$ increases at a fixed perturbation scale, $E_n$ remains nearly constant, while $\mathcal{F}_3$ decreases and $\mathcal{F}_4$ increases. These different trends reflect how the three functions respond to spatial variations. The error $E_n$ compares pseudospin directions pixel by pixel and depends on the local Stokes parameters and their perturbations. By contrast, $\mathcal{F}_3$ depends on the spatial derivatives of the pseudospin and its perturbation, while $\mathcal{F}_4$ further depends on the superposition of density errors from different points. These results show that changing the correlation length can produce different errors in topological observables, even though the Stokes perturbation scale is fixed. Therefore, $E_n$ and $\{\mathcal{F}_3,\mathcal{F}_4\}$ provide complementary evaluations of reconstruction quality in FSP.

        Therefore, we introduce a multi-stage optimization methodology for arbitrary FSP. The first part uses the PEF $\mathcal{F}_2$ to filter an optimal PSA configuration for Stokes reconstruction to reduce computational costs. This configuration then serves as the initial values for minimizing the TEF $\mathcal{F}_3$, with $\mathcal{F}_4$ and $E_n$ used as additional evaluating functions. For multiple incident textures (e.g, a full skyrmionic family), a weighted average of the evaluating functions over the sampled textures should be defined. This framework connects conventional Stokes polarimetry optimization to the topological observables. The specific FSP optimization cases and error verification will be discussed in Section 4.
        
\section{Configuration optimization}
    This section investigates whether the quality of topological reconstruction can be further improved after polarization optimization and evaluates the potential benefits of increasing the projection numbers in redundant measurements.
    
    \subsection{Multi-stage optimization}
        As mentioned above, the multi-stage optimization generates an optimal configuration $\mathbf{P}_L$ for polarization measurement, followed by an optimal configuration $\mathbf{P}_T$ for topological measurement. Considering a conventional bi-plate PSA with two adjustable azimuths $\{\beta_R,\beta_P\}$. When the projection numbers are $N$, both $\mathbf{P}_L$ and $\mathbf{P}_T$ are $N\times 2$ matrices in the form of $\{ \beta_{q}^{(1)},\beta_{p}^{(1)};\cdots; \beta_{q}^{(N)},\beta_{p}^{(N)} \}$. For the stage of polarization optimization, we combine Monte-Carlo sampling with the gradient descent method \cite{Wang2025}, sample 100 candidate configurations in the $2N-$dimensional parameter space first, and select the one with the lowest $\mathcal{F}_2$ as the starting configuration $\mathbf{P}_0$ (sampling is repeated if $\mathcal{F}_2<20$). Then, starting from $\mathbf{P}_0$, we optimize the PEF, $\mathcal{F}_2$, to a minimum to obtain the configuration $\mathbf{P}_L$.

        \begin{figure*}[htb]
    	\centering
        \includegraphics[width = 17.7cm]{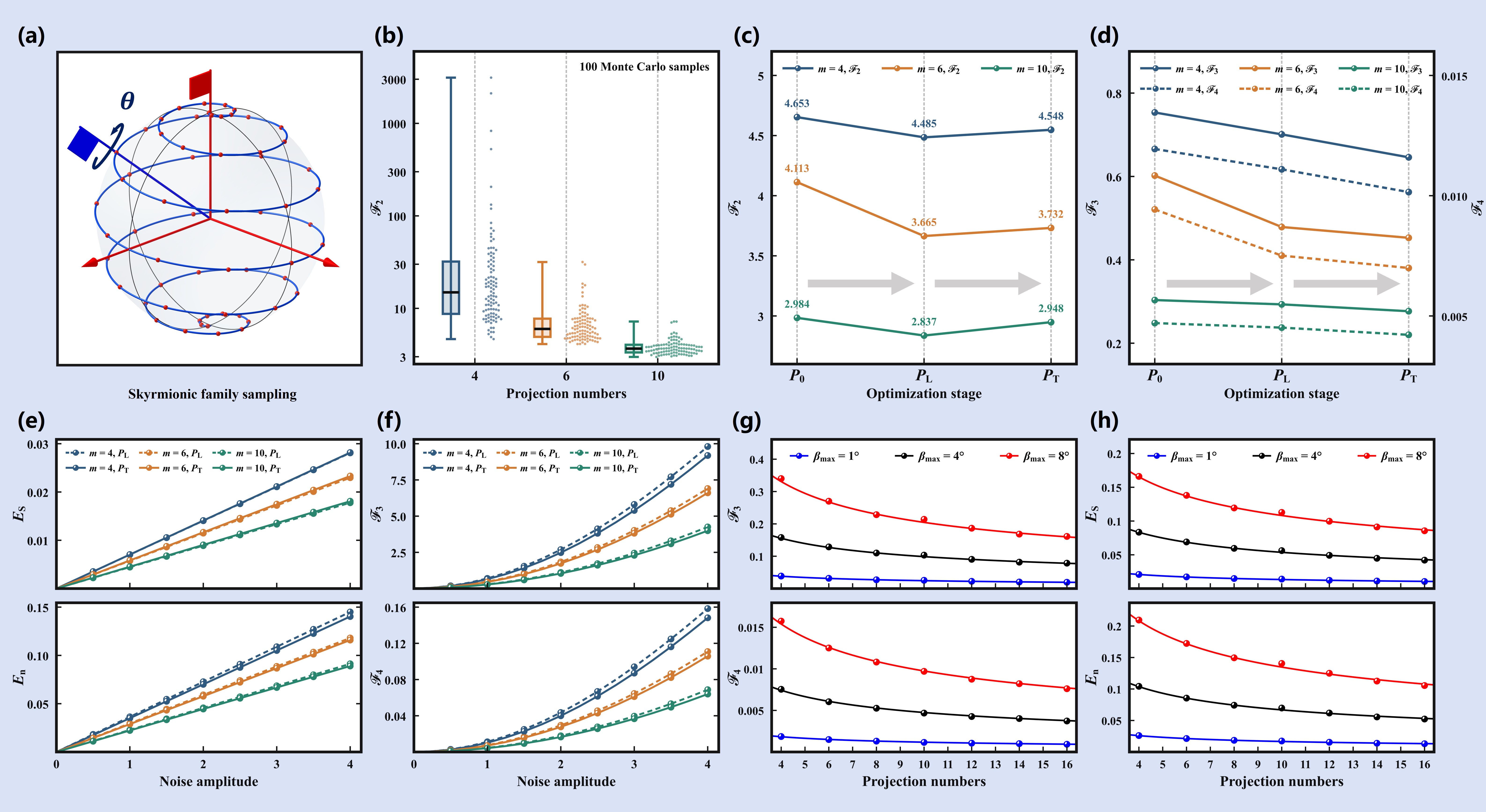}
           \caption{(a) Sampling of a full skyrmionic family, where 60 STF flagpole directions are sampled along a spiral, with four flag orientations at each direction, providing 240 sample points in total. (b) Monte-Carlo sampling of PSA configurations, where blue, orange, and green data points and box plots correspond to $N=4$, $6$, and $10$, respectively. (c) The PEF $\mathcal{F}_2$ and (d) the TEFs $\mathcal{F}_3$ (solid lines) and $\mathcal{F}_4$ (dashed lines) during multi-stage optimization for the three projection numbers. (e) and (f) The evaluating functions for the configuration $\mathbf{P}_L$ (dashed lines) and the topologically optimal configuration $\mathbf{P}_T$ (solid lines) with composite stochastic noise alone. (g) and (h) The evaluating functions at $\mathbf{P}_T$ as a function of projection number. Blue, black, and red data points correspond to $\beta_m=1^\circ$, $4^\circ$, and $8^\circ$, respectively. The curves are fits to the data points.}
           \label{fig:04}
           \vspace{-1.0\baselineskip}
        \end{figure*}
        
        Here, we consider the four-projection case with $N=4$ and two common redundant measurement cases with $N=6$ and $10$. FIG.4-(b) shows the distributions of $\mathcal{F}_2$ obtained by Monto-Carlo sampling, from which the initial configurations $\mathbf{P}_0$ are selected. These distributions verify that increasing the projection numbers makes it easier to find configurations with low noise amplification, which means measurement redundancy offers an advantage over four-projection measurements for high-precision FSP. Then, as shown in FIG.4-(c), about 100 iterations of gradient descent starting from $\mathbf{P}_0$ generate the optimal configuration $\mathbf{P}_L$ for polarization reconstruction. The $\mathcal{F}_2$ decreases further to 4.485, 3.665, and 2.837 for $N=4$, $6$, and $10$, respectively, approaching the corresponding theoretical lower bounds of this model. Then, we perform topological optimization by minimizing TEF, $\mathcal{F}_3$, through gradient descent to obtain $\mathbf{P}_T$. The TEFs are evaluated as weighted averages over a sampled full skyrmionic family \cite{Lara2009}, as shown in FIG.4-(a), with a unit level of composite stochastic noise. The optimization stops when the relative improvement $\Delta\mathcal{F}_3/\mathcal{F}_3$ is below $10^{-5}$ for 100 iterations or when the iteration count exceeds 30000. FIG.4-(c) and (d) show that although $\mathbf{P}_L$ is already optimal for polarization measurements, further topological optimization to $\mathbf{P}_T$ still reduces $\mathcal{F}_3$ and $\mathcal{F}_4$ by $5.4\sim7.8\%$ and $6.6\sim8.5\%$, respectively, with a small increase of $1.4\sim3.9\%$ in $\mathcal{F}_2$, which reflects that topological optimization can trade a small increase in Stokes parameter error for improved topological reconstruction.

        To test whether the benefits of topological optimization extend beyond the unit noise level, FIGs.4-(e) and (f) simulate the responses of configurations $\mathbf{P}_L$ and $\mathbf{P}_T$ to composite stochastic noise alone, where the composite noise source is modeled as $\delta \bm{I}(x,y)=\bm{n}_g+\bm{n}_p=\eta[(K-I)+\epsilon]$, with $K\sim\mathrm{Poisson}(I)$ and $\epsilon\sim\mathcal{N}(0,3^2)$. The amplitude factor $\eta=0$ corresponds to a noise-free measurement, while $\eta=1$ corresponds to the measurement with a unit noise level. For each incident texture, 128 complete simulated measurements are performed. In addition to $E_n$, $\mathcal{F}_3$, and $\mathcal{F}4$, the measurement results are also assessed by defining $E_S = [ \mathbf{E}\{ \iint_\Omega \|\bm{S}_R-\bm{S}_I\|^2 dxdy / \iint_\Omega \|\bm{S}_I\|^2 dxdy \} ]^{1/2}$, which directly quantifies the reconstruction error of the Stokes parameters. FIG.4-(e) shows that $E_S$ and $E_n$ increase approximately linearly with noise amplitude, whereas the topological errors $\mathcal{F}_3$ and $\mathcal{F}_4$ in FIG.4-(f) grow more rapidly. As $\eta$ increases from 2 to 4, $E_S$ and $E_n$ approximately double, while $\mathcal{F}_3$ and $\mathcal{F}_4$ increase fourfold. This behavior suggests that higher-order error contributions gradually become significant as the noise level continuously increases. Besides, The benefits of topological optimization are most obvious in the conventional four-projection case. Compared with $\mathbf{P}_L$, $\mathbf{P}_T$ reduces $\mathcal{F}_3$ and $\mathcal{F}_4$ by $6.3\%$ and $7.0\%$, respectively, while increasing $E_S$ by only $0.2\%$. As $N$ increases beyond four, the reconstruction errors at a given noise level decrease, and the performance difference between $\mathbf{P}_L$ and $\mathbf{P}_T$ becomes smaller. These results further highlight the value of measurement redundancy for high-precision FSP.

    \subsection{Projection redundancy}
        As introduced in FIG.2, DOT-FSP performs polarization projections sequentially in time, allowing a large number of projections $N$ at a low hardware cost, at the expense of temporal resolution. The Monte-Carlo sampling results and simulations with stochastic noise alone have shown that redundant projections improve measurement stability and reduce the effects of Error Source 2. Next, we investigate whether projection redundancy also reduces Error Source 1 caused by positional offsets of optical plates.

        With error source 2 excluded, we perform multi-stage optimization for each projection number $N=5,\ldots,16$, obtaining 12 topologically optimal configurations $\mathbf{P}_T$. In each projection, the QWR and LP are evaluated independent angular positional offset $\Delta\beta\in[-\beta_m,\beta_m]$, with $\beta_m=1^\circ$, $4^\circ$, or $8^\circ$. For each combination of $N$ and $\beta_m$, we generate 512 perturbed configuration matrices $\mathbf{P}_T'$ and simulate the corresponding measurements. Each $\mathbf{P}_T'$ contains $2N$ independent positional offset terms and produces a pseudospin field distortion shown in FIG.3-(a) and a topological density distortion described by Eq.(14). The final evaluating functions are calculated from these 512 measurements. In FIGs.4-(g) and (h), the solid lines are smooth fits to the 13 simulated data points in each series, including the standard case of $N=4$. All four evaluating functions, $E_S$, $E_n$, $\mathcal{F}_3$, and $\mathcal{F}_4$, show an overall decrease as the projection numbers increase. Increasing $N$ from 4 to 16 reduces these metrics by approximately half at all three offset levels. For instance, at $\beta_m=4^\circ$, $\mathcal{F}_3$ decreases from 0.1578 to 0.0785, and $\mathcal{F}_4$ decreases from 0.00754 to 0.00373. These significant improvements stem from the dispersion of error source 1 across a greater number of independent projections.

        The above results indicate that multi-stage optimization can further improve the reconstruction accuracy and stability of topological observables, even after the optimal configuration with polarization reconstruction has been established. Furthermore, in the case of projection redundancy, increasing projection numbers also provides a practical way to reduce the effects of both random matrix mismatch (Error Source 1) and stochastic noise (Error Source 2), which benefits the reconstruction of both polarization and topological observables.
        
\section{Discussion and conclusion}
    we define full-skyrmionic polarimetry (FSP) for measuring the complete skyrmionic characteristics (global and local) of skyrmionic beams, and develop its optimization framework from the reconstruction chain of both polarization and topological observables. Based on the stochastic error source models and polarization evaluating functions (PEF), $\{\mathcal{F}_1,\mathcal{F}_2\}$, we introduce two key topological evaluating functions (TEF), $\{\mathcal{F}_3,\mathcal{F}_4\}$, and an extra PEF, $E_n$, which provide effective metrics for quantitatively evaluating the measurement quality of FSPs (analogous to the condition number, CN).

    The presented theory and the numerical results in FIG.3 demonstrate that the topological protection of skyrmions does not ensure the immunity of topological observations against stochastic error sources. The stochastic error sources can distort the reconstructed local pseudospin field and topological density, introducing skyrmion number errors within a finite integration window. Besides, PEFs based on Stokes parameter perturbations are insufficient for evaluating the reconstructed topological observables. The normalization and spatial differentiation of the pseudospin field $\bm{n}(x,y)$ may lead to significantly different $\{\mathcal{F}_3,\mathcal{F}_4,E_n\}$ distributions, even when the Stokes parameter perturbations maintain the same scale. Therefore, we also develop a multi-stage optimization method, where polarization optimization is employed to define an optimal configuration $\mathbf{P}_L$ for polarization reconstruction first, and topological optimization is used to further determine another optimal configuration $\mathbf{P}_T$ for topological density reconstruction. As shown in FIG.4, this additional step improves topological reconstruction even after the configuration has reached the optimum for polarization observables.

    The simulations also demonstrate that for $N \ge 4$, projection redundancy offers a convenient way to improve measurement quality. Increasing the projection numbers reduces the effects of both error sources 1 and 2 on the reconstruction of full-skyrmionic observables. This stems from the averaging of error sources across redundant projections and is of critical significance for the division-of-time FSP (DOS-FSP), which allows for a large-scale increase in the number of redundant projections at a low cost to improve measurement quality. Future research could also incorporate automatic PSA control to advance the application of the FSP optimization framework for high-precision quasi-particle field measurements and potential skyrmion ellipsometry at weak-intensity levels.

\section*{Data Availability}
    The author confirms that the data supporting this study are available within the article.
    
\begin{acknowledgments}
    The author gratefully acknowledges the valuable discussions and suggestions provided by all co-authors during the development of this work. 
\end{acknowledgments}

\appendix

\section{Mueller calculus for MP-PSA}
    This section derives and expands Eq.(7), describing the polarization modulation of MP-PSA within the framework of Mueller calculus. Using the standard order of Stokes parameters $I \rightarrow Q \rightarrow U \rightarrow V$, the differential Mueller matrix of an independent polarization optical element (retarder, linear polarizer, beam splitter, or nanowire-grid) can be written as $\mathbf{O}\mathbf{M}_{\beta=0}\mathbf{O}^T$. Here, $\mathbf{M}_{\beta=0}$ is the Mueller matrix when the major axis is aligned with the X-axis. The orthogonal matrix $\mathbf{O}(\beta)$ describes the $\mathrm{SO}(3)$ rotation of the reference frame and is given by Eq.(A1). 
    \begin{equation}
        \begingroup
        % ===== 列间距：相邻两列的留白 = 2 × \arraycolsep =====
        \setlength{\arraycolsep}{7pt}  % 此处对应 12pt 的列间留白
        % ===== 行间距：基线上方高度 + 基线下方深度 =====
        \renewcommand{\arraystretch}{1}
        \setlength{\extrarowheight}{0pt}
        \setbox\strutbox=
        \hbox{
        \vrule height 9pt depth 6pt width 0pt
        }                            % 此处对应 13 + 5 = 18pt
        \mathbf{O}(\beta)=
        \begin{pmatrix}
            1 & 0 & 0 & 0 \\
            0 & \cos2\beta & -\sin2\beta & 0 \\
            0 & \sin2\beta & \cos2\beta & 0 \\
            0 & 0 & 0 & 1
        \end{pmatrix}
        \endgroup
    \end{equation}
    
    The same form also applies to other elements with equal or spatially varying principal axes, such as q-plates, J-plates, spiral phase plates, and acousto-optic or electro-optic crystals. The above coefficient 2 originates from a double covering of $\mathrm{SO}(3)$ by $\mathrm{SU}(2)$. For FIG.3-(b), the MP-PSA consists of phase retarders (PR) and linear polarizers (LP). The PR is described by the phase retardance $\delta$ between the polarization components along the major and minor axes, and the isotropic transmittance $T$. And the LP is described by the transmittances $T_1$ and $T_2$ for these two components, respectively. Their Mueller matrices at zero angular offset, $\mathbf{M}_{\beta=0}$, are provided by:
    \begin{equation}
        \begingroup
        % ===== 列间距：相邻两列的留白 = 2 × \arraycolsep =====
        \setlength{\arraycolsep}{6pt}  % 此处对应 12pt 的列间留白
        % ===== 行间距：基线上方高度 + 基线下方深度 =====
        \renewcommand{\arraystretch}{1}
        \setlength{\extrarowheight}{0pt}
        \setbox\strutbox=
        \hbox{
        \vrule height 9pt depth 6pt width 0pt
        }
        \begin{pmatrix}
            T & 0 & 0 & 0 \\
            0 & T & 0 & 0 \\
            0 & 0 & T\mathbf{C}_\delta & -T\mathbf{S}_\delta \\
            0 & 0 & T\mathbf{S}_\delta & T\mathbf{C}_\delta
        \end{pmatrix}
        \endgroup,\ 
        \begingroup
        % ===== 列间距：相邻两列的留白 = 2 × \arraycolsep =====
        \setlength{\arraycolsep}{7pt}  % 此处对应 12pt 的列间留白
        % ===== 行间距：基线上方高度 + 基线下方深度 =====
        \renewcommand{\arraystretch}{1}
        \setlength{\extrarowheight}{0pt}
        \setbox\strutbox=
        \hbox{
        \vrule height 9pt depth 6pt width 0pt
        }                            % 此处对应 13 + 5 = 18pt
        \begin{pmatrix}
            \mathrm{A} & \mathrm{B} & 0 & 0 \\
            \mathrm{B} & \mathrm{A} & 0 & 0 \\
            0 & 0 & \mathrm{C} & 0 \\
            0 & 0 & 0& \mathrm{C}
        \end{pmatrix}
        \endgroup
    \end{equation}    
    
    For simplification, we define the relationships $\mathrm{A}=(T_1+T_2)/2$, $\mathrm{B}=(T_1-T_2)/2$, $\mathrm{C}=\sqrt{T_1T_2}$, $\mathbf{s}_a=\sin{a}$, and $\mathbf{c}_a=\cos{a}$. We first consider the phase retarder, PR. Substituting the left-side matrix in Eq.(A2) into the decomposition expression $\mathbf{O}\mathbf{M}_{\beta=0}\mathbf{O}^T$ and simplifying provides an expanded form:
    \begin{equation}
        \begin{aligned}
            \begin{gathered}
                \mathbf{M}_R(\delta,\beta,T)=T\cdot \\
                \begingroup
                    % ===== 列间距：相邻两列的留白 = 2 × \arraycolsep =====
                    \setlength{\arraycolsep}{6pt}  % 此处对应 12pt 的列间留白
                    % ===== 行间距：基线上方高度 + 基线下方深度 =====
                    \renewcommand{\arraystretch}{1}
                    \setlength{\extrarowheight}{0pt}
                    \setbox\strutbox=
                    \hbox{
                    \vrule height 9pt depth 6pt width 0pt
                    }                            % 此处对应 13 + 5 = 18pt
                    \begin{pmatrix}
                        1 & 0 & 0 & 0 \\
                        0 & \mathbf{s}^2_{2\beta}\mathbf{c}_{\delta}+\mathbf{c}^2_{2\beta} & -\mathbf{s}_{\beta}\mathbf{c}_{2\beta}(\mathbf{c}_{\delta}-1) & \mathbf{s}_{2\beta}\mathbf{s}_\delta \\
                        0 & \mathbf{s}_{\beta}\mathbf{c}_{2\beta}(\mathbf{c}_{\delta}-1) & \mathbf{c}^2_{2\beta}\mathbf{c}_{\delta}+\mathbf{s}^2_{2\beta} & -\mathbf{c}_{2\beta}\mathbf{s}_{\delta} \\
                        0 & -\mathbf{s}_{2\beta}\mathbf{s}_\delta & \mathbf{c}_{2\beta}\mathbf{s}_{\delta} & \mathbf{c}_{\delta}
                    \end{pmatrix}
                \endgroup
            \end{gathered}
        \end{aligned}
    \end{equation}
    
    Then, we consider the linear polarizers, LP. Substituting the left-side matrix in Eq.(A2) into the decomposition expression similarly and simplifying provides another expanded form:
    \begin{equation}
        \begin{aligned}
            \begin{gathered}
                \mathbf{M}_P(\beta,T_1, T_2)= \\
                \begingroup
                    % ===== 列间距：相邻两列的留白 = 2 × \arraycolsep =====
                    \setlength{\arraycolsep}{6pt}  % 此处对应 12pt 的列间留白
                    % ===== 行间距：基线上方高度 + 基线下方深度 =====
                    \renewcommand{\arraystretch}{1}
                    \setlength{\extrarowheight}{0pt}
                    \setbox\strutbox=
                    \hbox{
                    \vrule height 9pt depth 5pt width 0pt
                    }                            % 此处对应 13 + 5 = 18pt
                    \begin{pmatrix}
                        \mathrm{A} & \mathrm{B}\mathbf{c}_{2\beta} 
                        & \mathrm{B}\mathbf{s}_{2\beta} & 0 \\
                        \mathrm{B}\mathbf{c}_{2\beta} & \mathrm{c}\mathbf{s}^2_{2\beta}+\mathrm{A}\mathbf{c}_{2\beta} & \mathbf{s}_{2\beta}\mathbf{c}_{2\beta}(\mathrm{A}-\mathrm{C}) & 0 \\
                        \mathrm{B}\mathbf{c}_{2\beta} & \mathbf{c}_{2\beta}\mathbf{c}_{2\beta}(\mathrm{A}-\mathrm{C}) & \mathrm{C}\mathbf{c}^2_{2\beta}+\mathrm{A}\mathbf{s}_{2\beta} & 0 \\
                        0 & 0 & 0 & 0
                    \end{pmatrix}
                \endgroup
            \end{gathered}
        \end{aligned}
    \end{equation}
    
    Following the Mueller matrix multiplication rule $\mathbf{M}=\mathbf{M}_m\cdots \mathbf{M}_1$, the total Mueller matrix $\mathbf{M}_i$ for the $i$th polarization projection of the MP-PSA takes the form provided in Eq.(7). Its explicit form is derived by substituting Eqs.(A3) and (A4) into Eq.(7).

\section{Topological density distortion}
    The scaling factor $g$ in Eq.(13) describes the distortion in topological density caused by random matrix mismatch (Error source 1). To theoretically derive it, we first consider the normalization of a nonzero polarization vector $\bm{v}=\{S_1,S_2,S_3\}$. The normalized pseudospin is $\bm{n}=\bm{v}/\|\bm{v}\|$. Differentiating this equation can provide:
    \begin{equation}
        \partial \bm{n} = \frac{\mathbf{I}_3-\bm{n}\bm{n}^T}{\|\bm{v}\|}\partial\bm{v}
    \end{equation}
    
    Here, $\partial\bm{v}$ denotes a small perturbation of the Stokes parameters. The operator $(\mathbf{I}_3-\bm{n}\bm{n}^{T})$ projects this perturbation onto the tangent plane of the pseudospin sphere in FIG.3-(a). Thus, only the component perpendicular to $\bm{n}$ changes the pseudospin direction, and its contribution is scaled by $1/\|\bm{v}\|$. When $\|\bm{v}_R\|$ approaches the noise level, the reconstructed pseudospin direction becomes highly sensitive to small perturbations. At $\bm{v}_R=0$, the normalized pseudospin $\bm{n}_R$ is undefined. In practice, the valid reconstruction region should be determined based on the signal-to-noise ratio of $\bm{v}_R$. This region should be used for the topological integrals of both the incident and reconstructed fields, as well as for TEF evaluation across different PSA configurations. 
    
    Next, we consider random matrix mismatch. For each complete projection-reconstruction, the matrix $\mathbf{C}$ and vector $\mathbf{b}$ on the left-side of Eq.(12) take random values that remain constant across the propagation plane. The reconstructed polarization vector is therefore $\bm{v}_R(x,y)=\mathbf{C}\bm{n}_I(x,y)+\mathbf{b}$, and the reconstructed pseudospin on the right-side of Eq.(12) is $\bm{n}_R=\bm{v}_R/\|\bm{v}_R\|$. By adjusting the notations and substituting the above relationships into Eq.(B1), the following expression can be derived:
    \begin{equation}
        \begin{aligned}
            \begin{gathered}
                \partial_{x,y}\bm{n}_R=\frac{\mathbf{I}_3-\bm{n}_R\bm{n}_R^T}{\|\bm{v}_R\|} \\
                =\frac{\partial_{x,y}\bm{v}_R}{\|\bm{v}_R\|}-
                \frac{\bm{v}_R\cdot(\bm{v}_R\partial_{x,y}\bm{v}_R)}{\|\bm{v}_R\|^3} \\
                =\frac{\partial_{x,y}\bm{v}_R}{\|\bm{v}_R\|}-
                \frac{\bm{v}_R\cdot\partial_{x,y}\|\bm{v}_R\|}{\|\bm{v}_R\|^2}
            \end{gathered}
        \end{aligned}
    \end{equation}

    Then, substituting Eq.(B2) into the definition of topological density in Eq.(5) and eliminating zero terms can further give: 
    \begin{equation}
        \begin{aligned}
            \begin{gathered}
                \rho_{sk}=\frac{1}{4\pi}\bm{n}_R\cdot
                (\partial_x\bm{n}_R \times \partial_y\bm{n}_R) \\ 
                =\frac{\bm{v}_R \cdot \left(\partial_x \bm{v}_R \times \partial_y \bm{v}_R\right)}{4\pi\|\bm{v}_R\|^3}
            \end{gathered}
        \end{aligned}
    \end{equation}
    
    The incident pseudospin field $\bm{n}_I(x,y)$ satisfies $\|\bm{n}_I\|=1$. Differentiating its squared norm with respect to $x$ and $y$ can give $\bm{n}_I\cdot\partial_x\bm{n}_I=\bm{n}_I\cdot\partial_y\bm{n}_I=0$. Therefore, both derivatives lie in the tangent plane of the reconstructed pseudospin sphere ($\bm{n}_I$) in FIG.3-(a). Their cross product is normal to this plane and can be written as a scalar multiple of $\bm{n}_I$. The coefficient is $\bm{n}_I\cdot(\partial_x\bm{n}_I\times\partial_y\bm{n}_I)$, which equals $4\pi q_{skI}$ by Eq.(9). Hence: 
    \begin{equation}
        \begin{aligned}
            \begin{gathered}
                \partial_x \bm{n}_I \times \partial_y \bm{n}_I=\left[ \bm{n}_I\cdot(\partial_x\bm{n}_I\times\partial_y\bm{n}_I)
                \right] \bm{n}_I \\
                =4\pi q_{skI} \bm{n}_I
            \end{gathered}
        \end{aligned}
    \end{equation}
    
    Since $\mathbf{C}$ and $\mathbf{b}$ are independent of the spatial coordinates $(x,y)$, the partial derivative of the vector $\bm{v}$ is given by $\partial_{x,y}\bm{v}=\mathbf{C}\partial_{x,y}\bm{n}$. Then, we introduce the scalar triple-product identity $(\mathbf{C}\bm{e}_1)\cdot[(\mathbf{C}\bm{e}_2)\times(\mathbf{C}\bm{e}_3)]=\det(\mathbf{C})[\bm{e}_1\cdot(\bm{e}_2\times\bm{e}_3)]$. Applying it together with Eq.(B4) to the numerator of Eq.(B3) can further provide:
    \begin{equation}
        \begin{aligned}
            \begin{gathered}
                \bm{v}_R(\partial_x\bm{v}_R \times \partial_y\bm{v}_R) \\ 
                =\mathbf{C}(\bm{n}_I+\mathbf{C}^{-1}\mathbf{b})
                \left[ (\mathbf{C}\partial_x\bm{n}_I) \times
                (\mathbf{C}\partial_y\bm{n}_I) 
                \right] \\
                =\mathbf{DET}(\mathbf{C})
                (\bm{n}_I+\mathbf{C}^{-1}\mathbf{b})
                (\partial_x\bm{n}_I\times \partial_y\bm{n}_I) \\
                =4\pi q_{skI} \mathbf{DET}(\mathbf{C})(1+
                \bm{n}_I^T\mathbf{C}^{-1}\mathbf{b})
            \end{gathered}
        \end{aligned}
    \end{equation}

    Finally, substituting Eq.(B5) into Eq.(B3) and using $\bm{v}_R=\mathbf{C}\bm{n}_I+\mathbf{b}$ can give the mapping from the incident topological density $q_{skI}(x,y)$ to the reconstructed topological density $q_{skR}(x,y)$ for a given matrix mismatch:
    \begin{equation}
        q_{skR}(x,y)=\frac{\mathbf{DET}(\mathbf{C})(1+\bm{n}_I^T\mathbf{C}^{-1}\mathbf{b})}
        {\|\mathbf{C}\bm{n}_I+\mathbf{b}\|^3}
        q_{skI}(x,y)
    \end{equation}

    Eq.(B6) provides the scaling factor $g(\bm{n}_I)$ in Eq.(13). The factor $\det(\mathbf{C})$ comes from the scaling of the scalar triple product. The term containing $\mathbf{b}$ accounts for the center shift of the transformed pseudospin sphere in FIG.3-(a), while the denominator arises from pseudospin renormalization. These three contributions together determine the distortion of the topological density.
    
% The \nocite command causes all entries in a bibliography to be printed out whether or not they are actually referenced in the text. This is appropriate for the sample file to show the different styles of references, but authors most likely will not want to use it.

%\nocite{*}

\bibliographystyle{apsrev4-2}
\bibliography{References}

\end{document}